\documentclass[aps,pre,nofootinbib,12pt]{revtex4}

\usepackage{amsmath}
\usepackage{amssymb}
\usepackage{epsfig}
\usepackage[latin1]{inputenc}
\usepackage{multirow}

\def\be{\begin{equation}}
\def\en{\end{equation}}

\def\RR{\rm \hbox{I\kern-.2em\hbox{R}}}
\def\NN{\rm \hbox{I\kern-.2em\hbox{N}}}
\def\ZZ{\rm {{Z}\kern-.28em{Z}}}
\def\CC{\rm \hbox{C\kern -.5em {\raise .32ex \hbox{$\scriptscriptstyle
|$}}\kern
-.22em{\raise .6ex \hbox{$\scriptscriptstyle |$}}\kern .4em}}

\def\<{\langle}
\def\>{\rangle}

\long\def\symbolfootnote[#1]#2{\begingroup%
\def\thefootnote{\fnsymbol{footnote}}\footnote[#1]{#2}\endgroup}

\begin{document}

\title{Short term forecasting of surface layer wind speed using a continuous cascade model}

\author{Rachel Ba\"{\i}le}
\email{baile@univ-corse.fr}
\affiliation{UMR CNRS 6134, Universit\'e de Corse, laboratoire de Vignola, rte des Sanguinaires, 20000 Ajaccio, France}

\author{Jean-Fran\c{c}ois Muzy}
\email{muzy@univ-corse.fr}
\affiliation{UMR CNRS 6134, Universit\'e de Corse, Campus Grossetti, 20250 Corte, France}

\author{Philippe Poggi}
\email{philippe.poggi@univ-corse.fr}
\affiliation{UMR CNRS 6134, Universit\'e de Corse, laboratoire de Vignola, rte des Sanguinaires, 20000 Ajaccio, France}

\begin{abstract}
This paper describes a statistical method for short-term forecasting of
surface layer wind velocity amplitude relying on the notion
of continuous cascades. Inspired by recent empirical findings that suggest the existence of some cascading process in the mesoscale range,
we consider that wind speed can be described by a seasonal component and a fluctuating part represented by a ``multifractal noise'' associated with a random cascade.
Performances of our model are tested on hourly wind speed series gathered at various locations in Corsica (France) and Netherlands.
The obtained results show a systematic improvement of the prediction as compared to reference models like persistence
or Artificial Neural Networks.
\end{abstract}

%\pacs{}

\maketitle

\section*{Keywords: Wind speed, random cascade, time series model, short term forecasting.}

\section{Introduction}

The fast growth of wind energy technology shows that more and more countries attach importance to this renewable resource.
However, the energy production is strongly dependent on the wind volatility and is consequently characterized by a large amount of
uncertainty. Reliable wind speed predictions are therefore necessary to optimize plants scheduling or to evaluate systems production.
For that purpose, many efforts have been spent for several years
by the scientific community in order to design faithful models that allow one to perform good forecasts.
As reviewed e.g. in \cite{anemos1}, there are mainly two families of approaches.
The ``physical'' models rely upon physical considerations leading to some atmospheric models
that provide a "numerical weather prediction" system.
For very short prediction horizons, one often prefers
"statistical" approaches that mainly consist in designing stochastic models or using methods of time series analysis, calibrated
on historical data or other explanatory variables (like the output of a physical model).
Within this framework, one can cite standard ARIMA modeling \cite{poggi,corotis2,DanChen}, models relying on Markov chains \cite{poggi2},
wavelet based methods \cite{kitagawa}, "black boxes" methods like advanced Recursive Least Squares or Artificial Neural Networks (ANN) \cite{karinio}.

The method we propose in this paper is based on recent empirical results according to which short time wind variations possess intermittent
statistical properties similar to those actually observed in fully developed isotropic turbulence \cite{Fri95}:
they are strongly non Gaussian and characterized by long range correlated (log-) amplitudes \cite{Muzy_Baile}.
These features have been shown to be the hallmark of random cascade processes.
We therefore propose to build a time series wind speed model involving a multifractal noise.
We aim at performing predictions of the wind intensity over horizons extending from 1 hour to 48 hours, using various data series
recorded at different sites located in Corsica (France) and Netherlands.
We then compare our results to those obtained with other common forecasting methods such as persistence, often considered as a reference, or an Artificial Neural Network.

The paper is structured as follows : in section \ref{data} are described the various time series used in this study. After a brief review of their main
linear properties (power spectrum, seasonality and correlations), we recall the observations of Muzy {\em et al.} \cite{Muzy_Baile} concerning
the statistics of wind variations amplitude. Section \ref{s_mo} is devoted to the definition of a simple stochastic multifractal model
for the wind velocity components relying on former observed features.
In section \ref{s_app}, we present results of the application of this model to short term predictions and comparison
to the aforementioned reference predictors. Conclusion and prospects are provided in section \ref{s_conc}.

\section{Description of the wind speed time series}
\label{data}

\subsection{Presentation of the data and basic statistical properties}
\label{sec_linear}

\begingroup
\squeezetable
 \begin{table}[h]
\begin{center}
 \begin{tabular}{|c|c|c|c|c|c|}
  \hline
   Location & Latitude & Longitude & Dates & Sampling freq. & Site\\
  \hline
   Vignola (Ajaccio) & $41^o56$'N & $8^o54$'E & 1998-2003 & 1 min & 70m, coastal, high hills\\
  \hline
   Ajaccio & $41^o55$'N & $8^o47$'E & \multirow{6}{*}{2002-2006} & \multirow{9}{*}{1 hour} & 5m, coastal, plain, airport \\
  \cline{1-3}\cline{6-6}
   Bastia & $42^o33$'N & $9^o29$'E & & & 10m, coastal, plain, airport \\
  \cline{1-3}\cline{6-6}
   Calvi & $42^o31$'N & $8^o47$'E & & & 57m, coastal, hills\\
  \cline{1-3}\cline{6-6}
   Conca & $41^o44$'N & $9^o20$'E & & & 225m, high hills\\
  \cline{1-3}\cline{6-6}
   Renno & $42^o11$'N & $8^o48$'E & & & 755m, mountains\\
  \cline{1-3}\cline{6-6}
   Sampolo & $41^o56$'N & $9^o07$'E & & & 850m, mountains\\
  \cline{1-4}\cline{6-6}
   Eindhoven & $51^o44$'N & $5^o41$'E & 1960-1999 & & 20m, plain\\
  \cline{1-4}\cline{6-6}
   Ijmuiden & $52^o46$'N & $4^o55$'E & 1956-2001 & & 4m, coastal, plain\\
  \cline{1-4}\cline{6-6}
   Schipol & $52^o33$'N & $4^o74$'E & 1951-2001 & & -4m, plain, airport \\
  \hline
\end{tabular}
\end{center}
\caption{Main features of the time series}
\label{table1}
\end{table}
\endgroup

The time series used in this paper are amplitude and direction of horizontal wind speeds recorded in Corsica (France) and Netherlands.
The first series called "Vignola", has been recorded (at 10 meters height) by the means of a cup anemometer, every minutes during 6 years (1998-2003) at our laboratory (Vignola) in Ajaccio, Corsica Island, France.
Other data sets consist in five years horizontal wind speed, determined every hour (10 minutes averages) for 6 sites in Corsica.
These data have been measured and collected by the french Meteorological Service of Climatology (Meteo-France)
using a cup anemometer and wind vane at $10$ meters above ground level.
For comparison purpose, we have also studied the wind data freely available from KNMI HYDRA PROJECT \cite{dataholland}.
These data represent series of hourly mean potential wind speeds recorded in 3 different sites in Netherlands.
Table \ref{table1} summarizes the main features of the sites.

\begin{figure}[h]
\begin{center}
\includegraphics[width=10cm]{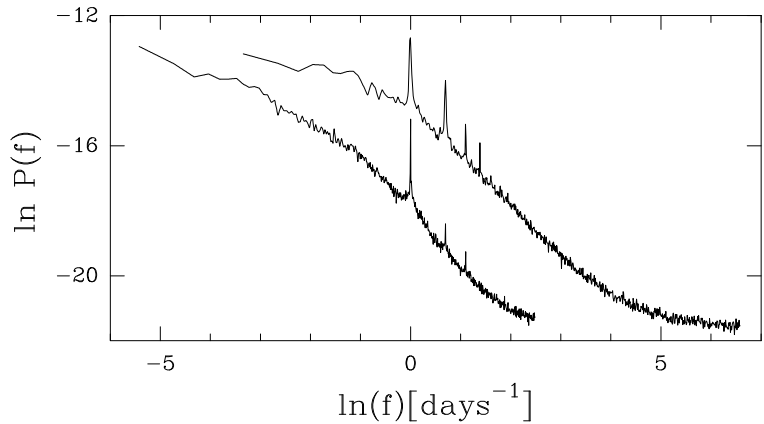}\\
\end{center}
\caption{Power spectrum density of wind intensity x-component series :  Vignola at the top and Eindhoven below, in log-log representation.}
\label{figSpectrum}
\end{figure}

In the sequel, $V(t)$ will denote the modulus of the velocity horizontal vector
while $V_x(t)$ and $V_y(t)$ will stand for its two components along arbitrary
orthogonal axes $x$ and $y$. We have by definition:
\[
   V(t) = \sqrt{V_x(t)^2+V_y(t)^2}
\]
Since our goal is to construct a parsimonious stochastic model of wind variations, we have chosen
to study $V_x$ and $V_y$ separately, a modeling of the wind dynamics in polar coordinates (modulus and direction)
would be cumbersome and more difficult to handle using Gaussian processes (see next section).
Moreover, since there is no well defined wind direction with a small "turbulent rate", components along
longitudinal and transverse directions are meaningless and we have preferred to focus on components along
arbitrary fixed directions.

\begin{figure}[h]
\begin{center}
\includegraphics[width=10cm]{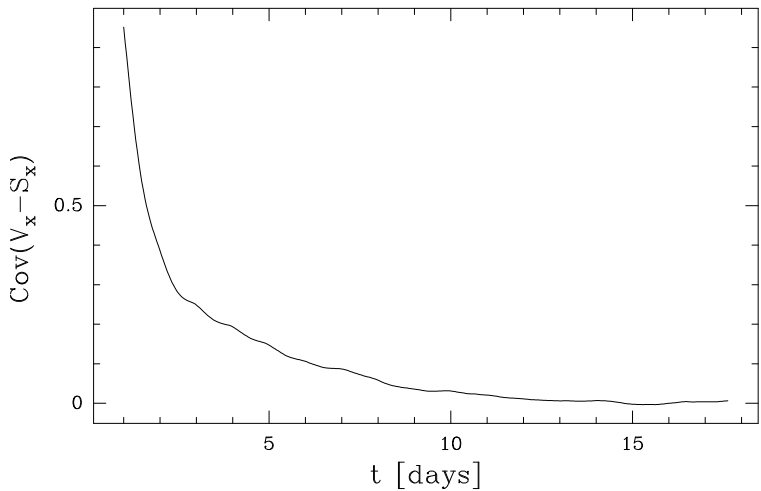}
\end{center}
\caption{Estimated covariance of deseasonalized $V_x$ component for Schipol (Netherlands).}
\label{figCorr1}
\end{figure}

The power spectrum is one of the most common tools
for analyzing random processes.
Since the pioneering work of Ven Der Hoven \cite{vdh,ot},
the shape of a typical atmospheric wind speed spectrum in the atmospheric
boundary layer is still matter of debate. It is relatively well admitted
that it possesses two regimes separated by low energy valley called the
``spectral gap'' located at frequencies around few
minutes. This gap separates the microscale regime, where turbulent
motions take place, from the mesoscale range.
In Fig. \ref{figSpectrum} are plotted, in log-log representation,
the power spectrum of $V_x$ wind component series corresponding to
Vignola and Eindhoven sites (for the sake of clarity, the graphs have been shifted by an arbitrary constant).
One sees that the Vignola spectrum (top curve) allows one to resolve higher
frequencies than the Eindhoven spectrum. In the former one, the beginning of
the spectral gap "plateau" can be clearly observed while the Eindhoven series goes down to smaller frequencies since it
covers a wider time period. The first striking feature of both spectra are
the main peaks associated with diurnal oscillations (see below). Up to the presence of these peaks,
both spectra can be represented by a decreasing function that connects the flat low
frequency behavior to the high frequency spectral gap.
The exact shape of the spectrum in the (intermediate) mesoscale range is unknown but it can be modeled by a power-law $P(f) \sim f^{-\beta}$ with an
exponent $\beta$ between $1.5$ and $2$.
One does not expect the same level of universality of the speed statistics as for turbulence
at mesoscale range and notably the value of the exponent can depend on local orographic, atmospheric conditions,...\cite{lmsa}.
Let us note that the spectrum associated with the $V_y$ velocity component behaves in a similar way.
The main power spectrum features can be alternatively observed through the behavior of the correlation function of the
velocity components. In Fig. \ref{figCorr1} is plotted the estimated
covariance of the de-seasonalized $V_x$ component of Schipol wind data as a function of the lag $\tau$
(see section \ref{s_mo} for the details about the way we process seasonal effects).
One sees that the correlation decreases quite slowly and the velocity remains correlated up to lags of few days.
Wind components auto-correlations and cross-correlations
can be easily described within the framework of linear time-series models.  ARMA like modeling \cite{BoxJen} have been widely
used to model many meteorological time series \cite{KatSka} like monthly precipitation \cite{DelKav}, annual streamflow \cite{CarMac} or
monthly drought index \cite{DavRap}. Many authors have also considered such time series models in order to account
for the fluctuations of the wind velocity amplitude or its components (see e.g., \cite{BlaDe,DanChen,KamJaf,poggi,McNew,McSpr1}).
However, since most of these approaches are faced to the non Gaussian nature of the wind fluctuations (see next section)
and the presence of seasonal effects, many of these models involve some non-linear "normalization" transformation and/or
a separate parametrization of each season \cite{BroKa84,Nfaoui,Baltsa,poggi}.
It results that, despite the simplicity of ARMA processes, the final models remain relatively hard to estimate and
far from being parsimonious. In this paper we choose to use (seasonal)
ARMA processes and account for the non-gaussian observed statistics through the nature of the noise term that will be given,
as explained in the next section, by a multifractal process.

\subsection{Non-linear statistical properties : non gaussian fluctuations and magnitude long-range correlations}

\begin{figure}
\begin{center}
\includegraphics[width=10cm]{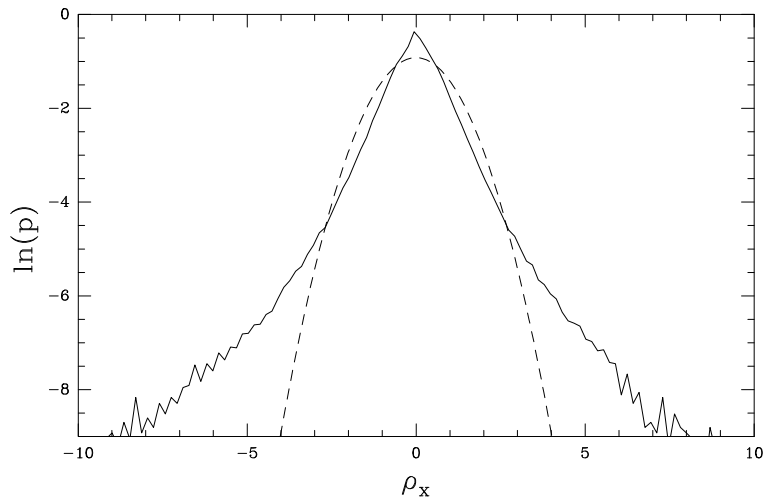}
\end{center}
\caption{Probability density function (pdf) of the error $\rho_{x}$ observed for an AR(1) model of
the $v_x$ component of Schipol wind series. The dashed line corresponds to a standardized normal law.}
\label{distribution}
\end{figure}

When referring to the non-Gaussian nature of wind speed statistics, one has to be precise since
velocity amplitudes are obviously not normally distributed. Indeed,
even in the case when $V_x$ and $V_y$
are Gaussian, the velocity modulus pdf is a Rayleigh distribution
(or a Rice distribution if the components have a non zero mean)
a particular case of the Weibull family. This is probably the main reason why Weibull is
the most commonly considered distribution in order to reproduce the pdf of wind amplitudes \cite{handbook}.
All the approaches that consist in directly trying to describe the stochastic dynamics of
the wind amplitude are faced to problems related to the non Gaussian nature of its statistics.
In particular, when one wants to account for the observed linear correlations, since the involved distributions
are not stable by aggregation (unlike Gaussian random variables), a control of both
persistence (correlations) and the nature of the statistics is very difficult \cite{Chou,McNew,McSpr2,GioUts}.
As mentioned previously, some authors have tried to reproduce the wind correlations, within AR Gaussian models, by using
a non-linear transformation of wind amplitudes in order to handle normal random variables \cite{DanChen,poggi}.
However, beyond the fact that these methods make strong assumptions on the nature of empirical laws,
they only account for the mono-variate distribution \symbolfootnote[2]{Indeed, even if marginal probability densities are
Gaussian, nothing guarantees that it is the case for the n-variate distributions} and
they are not stable as respect to aggregation, in the sense that
a change of the sampling period or the size of time averages would drastically modify
the parameters involved in the model.

As discussed in the previous section, a modeling of both components $V_x$ and $V_y$
allows one to reproduce the observed (partial) correlation functions within the
framework of ARMA models. However, even in the context, non Gaussian
statistics are observed: if one studies the distribution of the prediction errors of these models (or simply
the distribution of velocity components variations), it appears that their pdf are characterized by
stretched exponential tails very similar to the distribution of velocity
increments at small scales in fully developed turbulent flows \cite{Fri95}.
In Fig. \ref{distribution} is plotted the logarithm of the standardized pdf of the noise obtained
using an AR(1) process in order to model the variations of the Schipol series $V_x$ component
(the additive seasonal part has been removed). For comparison purpose we have
also plotted the parabola associated with a standardized Gaussian law.
It appears clearly that the pdf of noise fluctuations has a kurtosis very large
as compared to the normal law.

\begin{figure}
\begin{center}
\includegraphics[width=8cm]{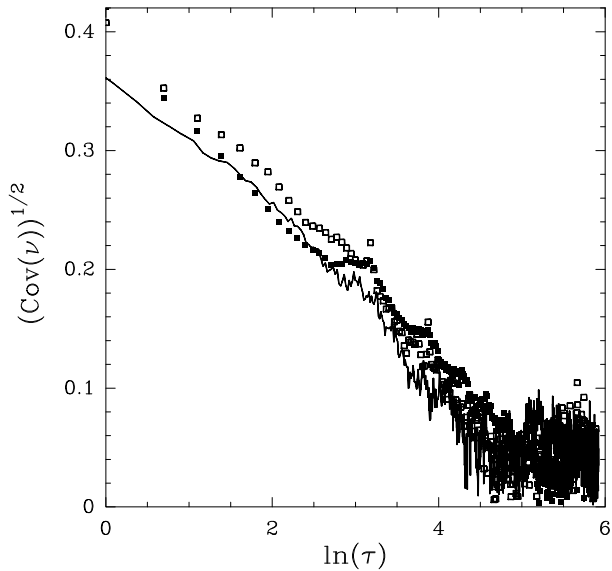}
\end{center}
\caption{Square root of wind magnitude covariance as a function
of the log of the lag (time units are hours).
The solid line corresponds to data from Vignola (20 minutes average).
The symbol $(\square)$ curve represents the average over the 7 Corsica sites, and the symbol $(\blacksquare)$ curve,
the mean of the 3 sites from Netherlands.}
\label{figCorr}
\end{figure}

Another striking feature of wind series is that the amplitude of the error noise is
long-range correlated. This is another property that has been observed in turbulence \cite{DelMuArn01,DelourPHD}.
More precisely, if one defines the local 'magnitude' as
$\nu(t)=\frac{1}{2}\ln\left(\rho_{x}(t)^{2}\right)$ or $\nu(t)=\frac{1}{2}\ln\left( \rho_{y}(t)^{2}\right) $ or
\be
\nu(t)=\frac{1}{2}\ln\left( \rho_{x}(t)^{2}+\rho_{y}(t)^{2}\right) ,
\label{eq_nu}
\en
where $\rho_x$ and $\rho_y$ are the noise terms associated with a linear prediction of $V_x$ and $V_y$
\symbolfootnote[3]{We would obtain the same results with $\nu_{s}(t)=\ln(\delta_{s}V_{x}^{2}+\delta_{s}V_{y}^{2})/2$
where $s$ is a small scale and $\delta_s F(t) = F(t+s)-F(t)$},
then the empirical covariance of $\nu$ can be fitted as:
\begin{equation}
\label{eq_cov}
\mbox{Cov} \left[\nu(t),\nu(t+\tau)\right] \simeq
\beta^2 \ln^2\left(\frac{\tau}{T}\right)  \; .
\end{equation}
By representing the square root covariance as a function of
the logarithm of the lag $\tau$ one should obtain a straight line.
This is illustrated in Fig. \ref{figCorr}, where the magnitude covariances estimated for
the sites in Corsica and Netherlands have been plotted (see caption).
It can be observed that the parameters $\beta^{2}$ (slope of the curves) and $T$ (time lag where correlations vanish)
are very close for all site data.
As explained in Ref. \cite{Muzy_Baile} and briefly reviewed in the next section,
these properties are intimately related to random cascade processes.

\subsection{Random cascade model for wind speeds}
\label{sect_cascade}

Discrete random multiplicative cascades were
originally introduced as models for
the energy cascade in fully developed turbulence.
%After the early works of Mandelbrot \cite{Man74a},
%a lot of mathematical studies have been devoted to random cascades \cite{KahPey76,Mol96}.
In the simplest case, these objects are positive fields (measures)
whose construction involves a recursive procedure along a dyadic tree:
the cascading process starts at a large ''integral''
time scale $T$ where the measure
is uniformly spread (meaning that the density is constant).
One then splits this interval in two
equal parts over which the densities are obtained by multiplying the 'father' density
by two (positive) i.i.d. random factors $W_1=e^{\kappa_1}$ and $W_2=e^{\kappa_2}$.
Each of these two sub-intervals is again cut in two equal parts and
the process is repeated infinitely. At construction step $n$,
the dyadic intervals have a size $T2^{-n}$ and their measure denoted $\sigma^2_{n}$ is simply:
$\sigma^2_{n}=\sigma_0^2 \prod W_{i}=2^{-n}e^{\sum\kappa_{i}}$,
where all the $W_{k}=e^{\kappa_{k}}$ are i.i.d such that $E(W)=1$.
If the random variables $\kappa$ are Gaussian,
then the corresponding model is log-normal and its scaling properties are easy
to control (see e.g., \cite{BacKozMuz06} and references therein for more details).
Let us notice that non positive fields like, for example, the velocity field in developed turbulence,
can be simply derived from the construction of the measure by considering that $\sigma^2(t)$ is the (stochastic) variance
of a Brownian motion (or another Gaussian process), i.e., $\delta_{\tau}X(t)= \sqrt{\sigma^2(\tau)} \varepsilon$ ,
where $\varepsilon$ is a Gaussian random noise.
% as described in previous section
Such "grid bounded" cascades, though simple, do not however provide a satisfying model for a
stationary physical process such as wind temporal variations.
Indeed, they are built on a fixed time interval $[0,T]$,
are not causal and not stationary. Moreover,
they involve an arbitrary fixed scale ratio (2 in the dyadic case).
Very recently, several constructions have been proposed to generalize
discrete cascades to stationary, causal and continuous processes \cite{BacKozMuz06,MuzDelBac00}.
We will not enter into details but if one calls the magnitude process $\omega(t)=\sum_i \kappa_i$,
then in the log-normal case, $\omega$ is a gaussian process characterized by its covariance function.
If one notices that the tree-like structure underlying the discrete construction implies a logarithmic
correlation function, then one can naturally define the log-normal continuous cascade
as follows \cite{MuzDelBac00,ArnMuzSor98}:
\be
\sigma^2_{s}(t)=e^{2 \omega_{s}(t)}
\label{eq_dx}
\en
where $\omega_{s}(t)$ is a stationary gaussian process of covariance defined by :
\be
\label{cascade}
\mbox{Cov}\left[\omega_{s}(t),\omega_{s}(t+\tau)\right]=\lambda^{2}\ln(\dfrac{T}{s+\tau}).
\en
Here $T$ and $\lambda^{2}$ are two parameters that correspond respectively
to the integral scale (correlation length analog to the time scale where cascading process starts) and the intermittency coefficient
(which quantifies the degree of burst occurrences in the process).
The parameter $s$ is a time sampling parameter that can be chosen arbitrary small (since the
weak limit $s \rightarrow 0$ of the process exists \cite{BacMuz03,BacKozMuz06}).
It can be proven that such a process is the continuous equivalent of discrete random cascades.
Therefore, according to this picture, a continuous cascade is nothing but a stochastic process
which magnitude, as defined by the logarithm of its variations,
has a covariance correlated as a logarithmic function.

In order to link these considerations with previous observed features for wind data,
let us remark that wind fluctuations at a fixed spatial location result
from two types of stochastic variations: first,
the spatial fluctuations at a fixed time (Eulerian) and then
the temporal fluctuations for a fixed fluid element (Lagrangian).
Since there is no strong mean velocity and Taylor
frozen hypothesis cannot be invoked, both Lagrangian and Eulerian variations have to be taken into account.
In ref. \cite{Cast02}, B. Castaing shows that
if one supposes a continuous cascade paradigm (Eq. \eqref{cascade})
for both Eulerian and Lagrangian fields, then the magnitude
correlation function at a fixed location should behave like
a squared logarithmic function:
\be
\mbox{Cov}\left[\omega_{s}(t),\omega_{s}(t+\tau)\right]=\beta^{2}\ln^{2}(\dfrac{T}{s+\tau})
\label{eq_cov}
\en
where the coefficient $\beta^{2}$ depends on both Lagrangian and Eulerian intermittency coefficients.
This is precisely the behavior that we observed in real data as reported in Fig. \ref{figCorr} of previous section
(see \cite{Muzy_Baile} for more details). Therefore,
the residual variance of errors $\rho_{x}(t)$ and $\rho_{y}(t)$ associated with linear models of $V_x(t)$ and $V_y(t)$
can be both defined as in Eq. \eqref{eq_dx} ($\rho_{x}(t)=e^{\omega_{s}(t)}\varepsilon_{x}(t)$
and $\rho_{x}(t)=e^{\omega_{s}(t)}\varepsilon_{y}(t)$) and:

\begin{equation}
%\nu(t)=ln(\rho_{x}(t)^{2}+\rho_{y}(t)^{2}) = 2Ms(t)+2\omega(t)+lnZ(t)
2 \nu(t)=\ln(\rho_{x}(t)^{2}+\rho_{y}(t)^{2}) = 2\omega(t)+\ln Z(t)
\label{eq_nu2}
\end{equation}
where $Z(t)=\varepsilon_{x}(t)^{2}+\varepsilon_{y}(t)^{2}$.\\

\section{Building the model}
\label{s_mo}
Let us now sum up all the reported empirical observations in
order to build a time series model of wind speed components $V_{x}(t)$ and $V_{y}(t)$.
According to previous considerations, the model will be
formulated as a seasonal auto-regressive process where
errors are given by a (seasonal) continuous cascade.

\subsection{Construction of the seasonal autoregressive part}
\label{sect_mod_seas}

It has been shown in section \ref{sec_linear} that $V_{x}(t)$ and $V_{y}(t)$
both contain additive seasonal components, i.e., can be written as:
\begin{equation}
\label{sdef}
V_{x,y}(t) = S_{x,y}(t) + V_{x,y}^S(t)
\end{equation}
where $S_{x,y}(t)$ represent the deterministic diurnal oscillations
and $V_{x,y}^S(t)$ the "de-seasonalized" velocity components.
Since the seasonality is caused by the variation of the sun position during the day, $S_{x,y}(t)$ are almost daily periodic functions, with
a period shape that changes according to the considered season in the year.
In order to determine this shape, we therefore have to perform a "local" estimation. For that purpose, we use a standard methodology described in \cite{Bovas}: each seasonal component
$S_{x}(t)$ and $S_{y}(t)$  (denoted as $S(t)$) is described by $m$ Fourier modes of period 1 $day$ ($D=24$ samples for hourly data):
\[
S(t) = \alpha_{0} + \sum_{k=1}^{m} \left[\eta_{1,k} \sin(\dfrac{2k\pi t}{D})+\eta_{2,k}\cos(\dfrac{2k\pi t}{D})\right]
\]
Because of the yearly variation of the seasonality, the coefficients $\{\eta_{i,k}\}_{i=1,2;k=1\ldots m}$ depend a priori
on the day $d$ and the local estimation simply
consists in using least squared method associated with a local exponential moving average:
\begin{equation}
\label{detalpha}
\left\{\eta_{i,k}\right\}_{i=1,2;k=1\ldots m}(d) = \mbox{argmin} \;\; \left\{ \sum_{yy=1}^{Y}\sum_{j} \psi^{|d-j|}\sum_{t=0}^{D-1}\left[V_{x,y}(yy,j,t)-S(t) \right]^2 \right\}.
\end{equation}
where $Y$ is the number of available years in the data series, $V_{x,y}(yy,j,t)$
represent the velocity component at year $yy$, day $j$ and 'hour' $t$. $\psi$ is an exponential discount factor
chosen so that $\frac{-1}{\ln(\psi)} \simeq 10$ days ($\psi = 0.9$).
We have used $D=24$ for hourly data and $m=3$.
We have checked that our results remain almost unchanged if one increases the number of harmonics.
Empirically, we have found that seasonal components represent $20$ to $35$ \% of the wind amplitude energy, except for 2 sites,
Ajaccio and Renno (Corsica) where they represent around $50$ \% of the total energy.

\begin{figure}
\begin{center}
\includegraphics[width=10cm]{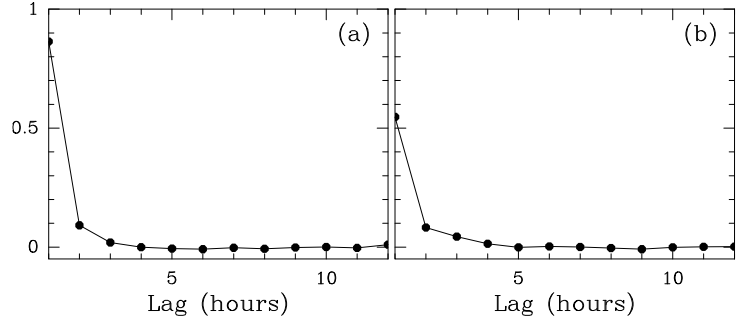}
\end{center}
\caption{PACF versus time lag : (a) corresponds to Schipol data and (b) to Ajaccio data.}
\label{pacf1h}
\end{figure}

In order to account for the linear correlations and cross-correlations of the stationary parts
$V_{x}^S(t)$ and $V_{y}^S(t)$, we have considered the class of
bi-variate ARMA processes.
The study of partial autocorrelation (PACF) and cross-correlation functions suggests that an AR of order 2 or 3 is appropriate
to fit the observations. This is illustrated in Fig. \ref{pacf1h}, where plots of PACF versus the lag
are reported for wind speed component $V_x^S$ of
Schipol and Ajaccio series. We have consistently observed that for all series,
the PACFs are close to zero value after lag 2. An AR(3) model should be more appropriate for some sites,
but accounting to higher order auto-regressive processes does not lead to any significant improvement of the results reported below.

The results of correlograms study can also be confirmed by other model selection procedures
like the Akaike Information Criterion (AIC).
The best choice of the AR order $p$ is the value that minimizes the following quantity :
\begin{equation}
AIC(p) = N \ln(\sigma^{2}_{\rho}(p))+2P,
\end{equation}
where $N$ is the length of each data series, $P$ is the number of estimated parameters and $\sigma^{2}_{\rho}$ is the variance of the residuals $\rho$.
We have studied this criterion for different sites and observed that
$AIC(p)$ decreases fast from $p=1$ to $p=2$ and slower at lags beyond $2$;
that confirms our previous results on the PACF and the choice of an AR(2)
model. Considering orders greater than 2 does not
improve the model forecasting performances.

Finally, we are lead to the following simple model for deseasonalized wind components:
\begin{equation}
\left\{
\begin{array}{ll}
V^{S}_{x}(t+1) = \sum_{k=0}^1\left(\gamma_{xx(k)}V^{S}_{x}(t-k)+\gamma_{xy(k)}V^{S}_{y}(t-k)\right)+\rho_{x}(t+1)\\
\\
V^{S}_{y}(t+1) = \sum_{k=0}^1\left(\gamma_{yy(k)}V^{S}_{y}(t-k)+\gamma_{yx(k)}V^{S}_{x}(t-k)\right)+\rho_{y}(t+1)\\
\end{array}
\right.
\label{eq_linear}
\end{equation}
where $\rho_{x,y}(t)$ represent
the noise terms which will be modeled as a log-normal continuous cascade,
(see next section), $\gamma_{xx(k)}$, $\gamma_{yy(k)}$, $\gamma_{xy}(k)$ and $\gamma_{yx}(k)$ ($k=1,2$) are
the AR coefficients.

Let us notice that the values of these coefficients strongly depend on the (arbitrary) choice of the reference direction defining $V_{x,y}$
and one cannot expect any universality or physical meaning in the precise value of each coefficient.
For instance, the coefficients estimated for the Schipol series are $\gamma_{xx(0)}=0.87$, $\gamma_{xx(1)}=0.09$, $\gamma_{xy(0)}=-0.05$ and $\gamma_{xy(1)}=0.04$
while the values we found for Ajaccio are $\gamma_{xx(0)}=0.56$, $\gamma_{xx(1)}=0.11$, $\gamma_{xy(0)}=0.06$
and $\gamma_{xy(1)}=-0.04$.

From a methodological point of view, we split the data in two parts :
the first part of each database (4 years for all Corsica sites, 20 for Ijmuiden,
30 for Eindhoven, 40 for Schipol)
was used as the "training period" or the "learning part". All the parameters of our model,
are determined using the learning part.
The remaining data allow us to evaluate the performance of each model as it will be seen later.

\subsection{Accounting for the cascade}
\label{mod_cascade}

As explained in section \ref{sect_cascade}, within the random cascade paradigm, the noise $\rho(t)$ can be written as :
\begin{equation}
\left\{
\begin{array}{ll}
\rho_{x}(t)=e^{\omega(t)+M_{s}(t)}\varepsilon_{x}(t)=e^{\Omega(t)}\varepsilon_{x}(t)\\
\rho_{y}(t)=e^{\omega(t)+M_{s}(t)}\varepsilon_{y}(t)=e^{\Omega(t)}\varepsilon_{y}(t)
\end{array}
\right.
\label{eq_ro}
\end{equation}
where $\varepsilon_{x,y}(t)$ are independent white Gaussian noises, $M_{s}(t)$ is a deterministic function that represents
a multiplicative seasonality of the noise amplitude and $\omega(t)$ is a zero mean stationary gaussian sequence independent of $\varepsilon(t)$,
which covariance is a squared log as described in section \ref{sect_cascade} (Eq. \eqref{eq_cov}).
In practice, one computes $ 2 \nu(t) = \ln(\rho_x^2(t)+\rho_y^2(t)) = 2 \omega(t) + 2 M_s(t) + \ln(\varepsilon_x^2(t) + \varepsilon_y^2(t))$
and since the mean and variance of $\ln Z(t) = \ln(\varepsilon_x^2(t) + \varepsilon_y^2(t))$ are known, one can obtain $M_s(t)$ along the same line
as we have estimated $S_{x,y}(t)$ (Eq. \eqref{detalpha}).
A generalized method of moments \cite{BacKozMuz06} applied to the sample covariance of $\nu(t)$ allows us to evaluate
the parameters $\beta^2$ and $T$ of Eq. \eqref{eq_cov}, defining the Gaussian process $\omega(t)$.

\section{Application to short term prediction}
\label{s_app}

\subsection{H-step forward prediction}
\label{hstep}

Our goal is to predict wind speed intensity $V(t) = \sqrt{V_x(t)^2+V_y(t)^2}$
at different horizons of time (from $1$ hour to $48$ hours).
Since the (conditional) law of the velocity modulus is not Gaussian,
the "best" prediction depends, in general, on the type of error one wants to minimize.
In theory, since the multifractal AR model we have introduced provides
the full conditional law of each velocity component, one should be able
to optimally solve any forecasting problem. For the sake of simplicity, we
will only estimate the conditional mean of $V$, denoted as $E(V|t)$ in the sequel,
that is the predictor which minimizes the mean square error.

Let $\hat{V}^S_{x,y}(t,h)$ (resp. $\hat{V}_{x,y}(t,h)$) be the best linear predictors of $V^S_{x,y}(t+h)$ (resp. $\hat{V}_{x,y}(t,h)$), at time $t$ and horizon $h$, i.e., from the definition of the model:
\begin{eqnarray*}
    \hat{V}^S_{x,y}(t,h) & = & E \left[V^S_{x,y}(t+h)|t \right] \\
    \hat{V}_{x,y}(t,h) & = & \hat{V}^S_{x,y}(t,h)+S_{x,y}(t+h)
\end{eqnarray*}
These predictors are easy to compute: since the linear part of our model reduces to a vector AR(2)
model (Eq. \eqref{eq_linear}), $h$ iterations of the model provide the linear coefficients. Indeed, Eq. \eqref{eq_linear} can be rewritten
in a vector form:
\begin{equation}
\label{vectormodel}
    \mathbf{V^S}(t+1) = \mathbf{\cal A} \mathbf{V^S}(t) + \mathbf{e}(t+1)
\end{equation}
where the vectors ${\mathbf V^S}(t)$ and $\mathbf{e}(t)$ are defined by:
\begin{equation}
\label{vect_Ve}
   {\mathbf V^S}(t) = \left(
\begin{array}{l}
V_x^S(t) \\
V_y^S(t) \\
V_x^S(t-1) \\
V_y^S(t-1)
\end{array}
\right)   , \;
\mathbf{e}(t) = \left(
\begin{array}{l}
\rho_{x}(t) \\
\rho_{y}(t) \\
0 \\
0
\end{array}
\right)   ,
\end{equation}
and the matrix $\mathbf{\cal A}$ reads:
\begin{equation}
 {\cal A} = \left(
 \begin{array}{cccc}
 \gamma_{xx}(0) & \gamma_{xy}(0) & \gamma_{xx}(1) & \gamma_{xy}(1) \\
 \gamma_{yx}(0) & \gamma_{yy}(0) & \gamma_{yx}(1) & \gamma_{yy}(1) \\
 1 & 0 & 0 & 0  \\
 0 & 1 & 0 & 0
\end{array}
\right) .
\end{equation}
When one considers an horizon $h$, the iteration of Eq. \eqref{vectormodel} gives:
\begin{equation}
\label{vectormodel2}
    \mathbf{V^S}(t+h) = \mathbf{\cal A}^h \mathbf{V^S}(t) + \sum_{k=0}^{h-1} \mathbf{\cal A}^k \mathbf{e}(t+h-k) = \mathbf{\cal A}^h \mathbf{V^S}(t) + \mathbf{e^{(h)}}(t+h).
\end{equation}
According to this representation, $\hat{V}^{S}_{x,y}(t,h)$ correspond to the first two components of $\mathbf{\cal A}^h \mathbf{V^S}(t)$.
From Eqs. \eqref{eq_ro} and \eqref{vect_Ve}, the components of the noise vector, in the r.h.s. of previous equation, can be written as:
\begin{equation}
\label{l1}
 e^{(h)}_{x,y}(t+h) = \sum_k a_k e^{\Omega(t+h-k)} \epsilon_{x,y}(t+h-k) \; ,
\end{equation}
where the constants $a_k$ can be deduced from the $\cal A$ coefficients. Moreover, by considering,
as shown in ref. \cite{BacKozMuz09}, $\epsilon(t) e^{\Omega(t)}$
quasi-stable as respect to linear combinations,
we have:
\begin{equation}
\label{l2}
 e^{(h)}_{x,y}(t+h) \operatornamewithlimits{=}_{law} e^{\Omega^{(h)}(t+h)} \epsilon^{(h)}_{x,y}(t+h)
\end{equation}
where $\epsilon^{(h)}$ is a standardized Gaussian noise and $\Omega^{(h)}$ is also Gaussian, at fixed $h$, with the same covariance as $\Omega(t)$ for lags
greater than $h$ (Eq. \eqref{eq_cov}).
Eqs. \eqref{vectormodel2} and \eqref{l2} show that the model conserves the same shape for all prediction horizons:
\begin{equation}
\label{model_h}
   V_{x,y}^{S}(t+h) = \hat{V}^{S}_{x,y}(t,h) + e^{\Omega^{(h)}(t+h)} \epsilon^{(h)}_{x,y}(t+h).
\end{equation}
This property is of great practical interest because, whatever the horizon $h$, at fixed value of $\Omega^{(h)}(t+h)$,
the law of the velocity modulus $V(t+h)$ is a Rice distribution \cite{Rice} of parameters $r = \sqrt{\hat{V}_x^2(t+h)+\hat{V}_y^2(t+h)}$ and $\sigma^2=e^{2\Omega^{(h)}(t+h)}$.
More specifically, let $M_R(r,\sigma^2)$ be the mean value of a Rice distribution, i.e.,
\begin{equation}
  M_R(r,\sigma^2) = \sigma \sqrt{\frac{\pi}{2}} L_{1/2}\left(-\frac{r^2}{2 \sigma^2}\right)
\end{equation}
(where $L_{1/2}(x)$ is the order $1/2$ Laguerre polynomial), and $P_h(\Omega|t)$ the conditional Gaussian law of
$\Omega^{(h)}(t+h)$. The conditional velocity value at horizon $h$ is then:
\begin{equation}
   E\left(V(t+h)|t\right)  = \int P_h(\Omega|t) M_{R}(r,e^{2\Omega}|t) d\Omega.
\end{equation}
This quantity can be evaluated numerically by a Gaussian quadrature approximation of
the Gaussian integral \cite{PreTeuVetFla88}.
The conditional law of $\Omega^{(h)}(t+h)$ is a normal law which mean, $\hat{\Omega}^{(h)}(t+h)$, and variance, $s_{\Omega}^{(h)}(t+h)$,
can be computed using the known mean and covariance of $\Omega^{(h)}$.
$\hat{\Omega}^{(h)}(t+h)$ is nothing but the best linear predictor of $\Omega(t+h)$ at time $t$ and horizon $h$, i.e.:
\begin{equation}
\hat{\Omega}^{(h)}(t+h)= M_S^{(h)}(t+h)+ \sum_{k=0}^{T-1}\alpha_{k}\omega^{(h)}(t-k),
\end{equation}
where the filter size $T$ and the coefficients $\alpha_k$ are obtained from the shape of the covariance function
of $\omega^{(h)}$ (Eq. \eqref{eq_cov}). If one denotes
$C^{(h)}_{ij} = \mbox{Cov}\left[\omega^{(h)}(t),\omega^{(h)}(t+|j-i|)\right]$
and $\zeta^{(h)}_{k} = \mbox{Cov}\left[\omega^{(h)}(t),\omega^{(h)}(t+k+h)\right]$,
then
\begin{equation}
 \alpha_k = \sum_j \left[C^{(h)}_{kj}\right]^{-1} \zeta^{(h)}_j.
\end{equation}

Let us end this section by noticing that the alternative predictor
\begin{equation}
 {\hat V}(t+h) = \sqrt{ E\left(V^2(t+h) | t \right)} \;,
\end{equation}
which, after a little algebra, reduces to
 \begin{equation}
\hat{V}(t+h) = \sqrt{\hat{V}_{x}(t+h)^{2}+\hat{V}_{y}(t+h)^{2}+2e^{2\hat{\Omega}^{(h)}(t+h)+2s_{\Omega}^{(h)}(t+h)}}
\label{predic_V}
\end{equation}
provides performances relatively close to the former ``Rice'' predictor.

\subsection{Forecasting performances of our wind model}
\label{s_mod}
We present in this section the forecasting performances of the
previously defined model as compared to standard models like persistence,
a reference model introduced by Nielsen \textit{et al}. \cite{nielsen}
and a simple Artificial Neural Network (ANN).
The parameters of these two latter models are estimated over
the previously defined "learning part" of each data series (see section \ref{sect_mod_seas}).
Models comparison are made using two different mean error measurements.
\subsubsection{Reference models}
\label{ss_ref}

As explained in \cite{anemos1}, simple techniques are often
used as references within the wind power forecasting community. Let us briefly
describe the 3 main models we considered for performance comparison purpose.

$\bullet$ \textit{Persistence}

This model is the most commonly used reference predictor.
According to Giebel \cite{giebel},
for short prediction horizons (from few minutes to hours),
this model is the benchmark all other prediction models have to beat.
It consists in a simple martingale hypothesis according to which
future wind speed at horizon $h$ will be the same as the present observed value :

\be
\hat{V}(t+h|t) = V(t).
\en

$\bullet$ \textit{Merge of persistence and global average}

Nielsen \textit{et al}. \cite{nielsen} propose to use a linear combination of
the persistence predictor and the global average to improve the previous persistence prediction:
\be
\hat{V}(t+h|t) = aV(t) + (1-a)\overline{V},
\en
where $a$ is the correlation coefficient between $V(t)$ and $V(t+h)$ and $\overline{V}$ is
the mean velocity. $\overline{V}$ and $a$ can be determined using data up to time $t$ or using
the chosen training period of each database.
Let us note that $V-\overline{V}$ can be identified as an AR(1) predictor.

$\bullet$ \textit{Artificial Neural Network (ANN) model}

Artifical neural networks are commonly used as "black boxes" prediction tools in many areas.
Notably, there is a wide literature on their interest in wind speed forecasting (see e.g.
\cite{anemos1} and references therein).
We have designed this method using the ANN toolbox of MATLAB,
with the collaboration of Philippe Lauret, as introduced in \cite{lauret}.
We have chosen the most popular form of NN called multilayer perceptron (MLP) structure.
The MLP structure consists of an input layer,
one or several hidden layers and an output layer.
In our case, the input vector is given by the previous
observed values of the wind speed and the output vector consists
of only one output, which is the corresponding forecast at horizon $h$.
Best results are here obtained with 30 input neurons and one hidden layer,
characterized by 5 non-linear units (or neurons).
The non-linear function associated with each unit is usually a tangent hyperbolic
function $f(x)=\tanh(x)$.
Therefore, a NN with $N_i=30$ inputs, $N_h=5$ hidden neurons
and a single linear output unit defines a non-linear
parameterized mapping from an input $x$ to an output $y$, given by :
\begin{equation}
y = y(x;w) = \sum_{j=0}^{N_h}\left[ w_{j}f\left( \sum_{i=0}^{N_i}w_{ji}.x_{i}\right) \right]
\end{equation}
where $w_{j}$ are the weight applied on each hidden neuron and $w_{ji}$ ones applied on each input data.
These NN parameters $w$ are estimated during a learning phase. It
consists in adjusting $w$ so as to minimize an error
function which is usually the sum of squares error between measured data and network output (see next section).
For that purpose, several iterations are necessary (we have observed that 30 are sufficient).

\subsubsection{Estimation of forecasting accuracy}
\label{ss_mod}

%According to \cite{anemos2}, "
Errors frequently used to compare various prediction methods
are the Root Mean Square Error (RMSE),
the Mean Absolute Error (MAE), the Mean Error (ME),
histograms of the frequency distribution of the error or the correlation function \cite{anemos1}.
We have chosen to employ the most common of them, i.e., the RMSE
and the MAE, in order to evaluate the relative performances of each model.
These errors are given as percent of the mean of wind speed at each site.
If $V(t)$ is the observed wind speed at time $t$ and
$\hat{V}(t)$ the corresponding forecast, these errors are defined as follows:

$\bullet$ The normalized mean absolute error (nMAE) is simply defined as:
\be
nMAE = \frac{1}{\overline{V}}\frac{1}{n}\sum_{t=1}^n \mid V(t)-\hat{V}(t)\mid,
\en
where $n$ is the number of periods of time and ${\overline{V}}$ is the mean velocity amplitude
over the testing period.

$\bullet$ The normalized root mean square error (nRMSE), which
gives more weight to largest errors, reads:
\be
nRMSE = \frac{1}{\overline{V}}\sqrt{MSE}
\en
with
\be
MSE = \frac{1}{n}\sum_{t=1}^n (V(t)-\hat{V}(t))^{2}
\en

\subsubsection{Results}
\label{ss_res}

\begin{table}[h]
\label{tab_rmse}
\begin{center}
 \begin{tabular}{|c||c|c|c||c|}
  \hline Location & Pers. & Pers+Mean & RNA & Mult. mod.\\
  \hline
  Vignola & 42.7 & 39.8 & 38.4 & \textbf{37.6}\\
  \hline
   Ajaccio & 40.4 & 36.6 & 34.8 & \textbf{33.8}\\
  \hline
   Bastia & 44.9  & 42.1 & 40.4 & \textbf{40.2}\\
  \hline
   Calvi & 40.2 & 38.4 & \textbf{35.7} & 36.0\\
  \hline
   Conca & 49.7 & 47.4 & \textbf{46.0} & \textbf{46.0}\\
  \hline
   Renno & 44.1 & 40.5 & 39.1 & \textbf{37.6}\\
  \hline
   Sampolo & 54.4 & 51.6 & 48.3 & \textbf{47.9}\\
  \hline
   Ijmuiden & 13.6 & \textbf{13.5} & \textbf{13.5} & 13.6\\
  \hline
   Schipol & 17.5 & 17.3 & 17.1 & \textbf{16.9}\\
  \hline
   Eindhoven & 20.3 & 20.0 & 19.8 & \textbf{19.7}\\
  \hline
\end{tabular}
\end{center}
\caption{nRMSE (\%) of each site at one hour horizon. Best results are indicated using bold faces.}
\end{table}

\begin{table}[h]
\label{table_rmse6}
\begin{center}
 \begin{tabular}{|c||c|c|c||c|}
  \hline Location & Pers. & Pers+Mean & RNA & Mult. mod.\\
  \hline
  Vignola & 70.3 & 56.2 & 53.5 & \textbf{51.2}\\
  \hline
   Ajaccio & 66.6 & 48 & 43.1 & \textbf{41.4}\\
  \hline
   Bastia & 77.1 & 61.8 & 57.5 & \textbf{55.3}\\
  \hline
   Calvi & 66.2 & 57.7 & 54.4 & \textbf{52.0}\\
  \hline
   Conca & 78.7 & 69.1 & 66.7 & \textbf{66.4}\\
  \hline
   Renno & 71.3 & 54.9 & 52.4 & \textbf{49.6}\\
  \hline
   Sampolo & 101.9 & 81.8 & 69.4 & \textbf{65.4}\\
  \hline
   Ijmuiden & 33.5 & 31.6 & 31.4 & \textbf{31.3}\\
  \hline
   Schipol & 43.6 & 40.2 & 38.7 & \textbf{36.7}\\
  \hline
   Eindhoven & 47.6 & 43.5 & 41.5 & \textbf{39.5}\\
  \hline
\end{tabular}
\end{center}
\caption{nRMSE (\%) of each site at 6 hours horizon. Best results are indicated using bold faces.}
\end{table}

\begin{figure}
\begin{center}
\includegraphics[width=8cm,angle=270]{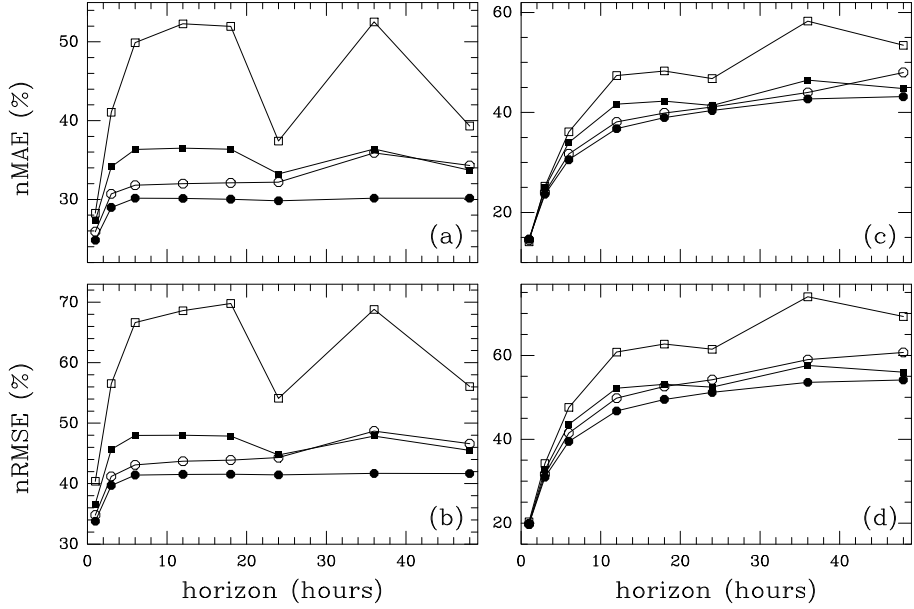}
\end{center}
\caption{Evolution of RMSE and MAE values for different models
depending on the horizon (1 hour to 48 hours).
Figures (a) and (b) illustrate these evolutions for Ajaccio (Corsica),
(c) and (d) correspond to Eindhoven in Netherlands.
For each case, symbol $(\square)$ curve
represents the persistence model's results, $(\blacksquare)$ curve,
the merge of persistence and global average, $(\circ)$ curve,
the ANN model and $(\bullet)$ curve, the multifractal model.}
\label{figRMSE}
\end{figure}

In Fig. \ref{figRMSE} the nMAE and nRMSE associated with each model prediction
are represented, at various forecasting horizons, for 2 sets of data
(Ajaccio in Corsica and Eindhoven in Netherlands).
For the $1$ hour horizon,
as it can also be observed for the nRMSE in table \ref{tab_rmse}, the performances
obtained with the cascade model are slightly better than those obtained
with concurrent models (average improvement of respectively 1 and 10 percent as compared to ANN and persistence).
When the horizon increases, the performances of each model decrease,
but the relative accuracy of our model becomes more and more significant.
This is confirmed in table \ref{table_rmse6} where are reported the nRMSE at 6 hours horizon
for all the data series (average improvement of respectively 4 and 26 percent as compared to ANN and persistence).

\begin{figure}
\begin{center}
\includegraphics[width=8cm]{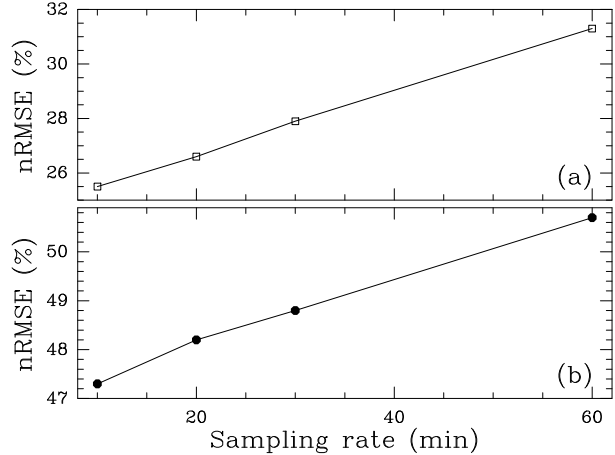}
\end{center}
\caption{nRMSE values evolution of the hourly mean Vignola series forecast using multifractal model, depending on the database sampling rate (10 minutes to one hour). Figure (a) illustrates this evolution for one hour horizon forecast whereas (b) corresponds to 6 hours horizon forecast.}
\label{figRMSE_vig}
\end{figure}

We have also evaluated the models performances when
one increases the sampling frequency of the data used to compute the prediction,
for some fixed time scale and horizon.
The data set gathered at Vignola, sampled at 1 minute rate,
allows us to compare the forecasts of hourly mean velocity, 1 and 6
hours ahead, by using velocity data at different sampling rates:
10 minutes, 20 minutes, 30 minutes and 1 hour.
In Fig. \ref{figRMSE_vig} are reported the prediction errors
of the cascade model as a function of the sampling rate for fixed
horizon and averaging time scale. One clearly sees a systematic improvement
of the accuracy as one uses better resolved input data:
the finer the sampling rate, the better the forecast. Similar
improvements can be observed with others models.
This result highlights the importance of having
high frequency data to enhance the forecast quality.

\section{Conclusion}
\label{s_conc}
In this paper, we have addressed the problem of short term wind speed forecasting
using a simple autoregressive seasonal model involving multifractal fluctuations.
This model relies on "universal"
empirical observations showing that high frequency velocity components
variations have long range correlated amplitudes \cite{Muzy_Baile}.
Our model is relatively parsimonious and accounts for the wind properties over all time scales.
It has been applied to forecast hourly wind speed data up to two days (48h) ahead.
The obtained results show that the proposed method is more accurate than
standard reference models. Let us notice that our approach can be
improved by considering, for instance, its natural
multivariate generalization.
This may allow us to describe the joint wind variations at different
locations. Let us also mention that unlike 'black boxes' approaches,
our time series cascade model is able to provide unconditional and
conditional velocity probability distributions and therefore address
many questions related to resource assessment or risk management. This
problem will be the scope of a further study.

%\begin{
\section*{Acknowledgments}
%\phantomsection
We would like to thank Philippe Lauret and the LPBS laboratory of
Reunion University for receiving one of us and for his helpful advice
in designing ANN forecasting tools.
We would also like to thank Meteo-France for the access
to their data and the Royal Netherlands Meteorological Institute for
providing free wind data online.
%\end{acknowledgments}

\section*{References}

%\bibliographystyle{elsarticle-num}
%\bibliographystyle{is-unsrt}
%\bibliographystyle{phiaea}
%\bibliography{BibWind}

\newpage

\end{document}